\documentclass[sigconf]{acmart}
\usepackage[utf8]{inputenc} 
\usepackage[T1]{fontenc}    
\usepackage{url}            
\usepackage{booktabs}       
\usepackage{amsfonts}       
\usepackage{nicefrac}       
\usepackage{microtype}      
\usepackage{multirow}
\usepackage{enumitem}
\usepackage{amsmath,amssymb,amsfonts}
\usepackage{graphicx}
\usepackage{caption}
\usepackage{listings}
\usepackage{textcomp}
\usepackage{xcolor}
\usepackage{subcaption}
\usepackage{tcolorbox}
\usepackage{xspace}
\usepackage{todonotes}
\usepackage{flushend}

\newtheorem{Def}{Definition}
\newcommand{\lenet}{\hbox{\emph{LeNet-5}}\xspace}
\newcommand{\nin}{\hbox{\emph{NIN}}\xspace}
\newcommand{\res}{\hbox{\emph{ResNet-20}}\xspace}
\newcommand{\mobile}{\hbox{\emph{MobileNet}}\xspace}

\newcommand{\eg}{\hbox{\emph{e.g.}}\xspace}
\newcommand{\ie}{\hbox{\emph{i.e.}}\xspace}
\newcommand{\wrt}{\hbox{\emph{w.r.t.}}\xspace}

\newcommand{\tool}{\emph{\textit{KuK}}\xspace}

\copyrightyear{2020}
\acmYear{2020}
\setcopyright{acmcopyright}
\acmConference[ICSE '20]{42nd International Conference on Software Engineering}{May 23--29, 2020}{Seoul, Republic of Korea}
\acmBooktitle{42nd International Conference on Software Engineering (ICSE '20), May 23--29, 2020, Seoul, Republic of Korea}
\acmPrice{15.00}
\acmDOI{10.1145/3377811.3380368}
\acmISBN{978-1-4503-7121-6/20/05}

\begin{document}

\title[Characterizing Adversarial Defects of DL Software from the Lens of Uncertainty]{
	Towards Characterizing Adversarial Defects
	of Deep Learning Software from the Lens of Uncertainty}

\author{Xiyue Zhang}\authornote{Co-first authors}
\affiliation{Peking University, \\ China}

\author{Xiaofei Xie}\authornotemark[1]
\affiliation{Nanyang Technological University, Singapore}

\author{Lei Ma}\authornote{Lei Ma (malei@ait.kyushu-u.ac.jp) and Meng Sun (sunm@pku.edu.cn) are the corresponding authors.}
\affiliation{Kyushu University, \\ Japan}

\author{Xiaoning Du}
\affiliation{Nanyang Technological University, Singapore}

\author{Qiang Hu}
\affiliation{Kyushu University, \\ Japan}

\author{Yang Liu}
\affiliation{Nanyang Technological University, Singapore}

\author{Jianjun Zhao}
\affiliation{Kyushu University, \\ Japan}

\author{Meng Sun}\authornotemark[2]
\affiliation{Peking University, \\ China}

\renewcommand{\shortauthors}{Xiyue Zhang, Xiaofei Xie, Lei Ma and et al.}

\begin{abstract}

Over the past decade, deep learning (DL) has been successfully applied to many industrial domain-specific tasks.
However, the current state-of-the-art DL software still suffers from quality issues, which raises great concern especially in the context of safety- and security-critical scenarios. Adversarial examples (AEs) represent a typical and important type of defects needed to be urgently addressed, on which a DL software makes incorrect decisions. Such defects occur through either intentional attack or physical-world noise perceived by input sensors, potentially hindering further industry deployment.
The intrinsic uncertainty nature of deep learning decisions can be a fundamental reason for its incorrect behavior. 
Although some testing, adversarial attack and defense techniques have been recently proposed, it still lacks a systematic study to uncover the relationship between AEs and DL uncertainty. 

In this paper, we conduct a large-scale study towards bridging this gap. We first investigate the capability of multiple uncertainty metrics in differentiating benign examples (BEs) and AEs, which enables to characterize the uncertainty patterns of input data.
Then, we identify and categorize the uncertainty patterns of BEs and AEs, and find that while BEs and AEs generated by existing methods do follow common uncertainty patterns, some other uncertainty patterns are largely missed. 
Based on this, we propose an automated testing technique to generate multiple types of uncommon AEs and BEs that are largely missed by existing techniques.
Our further evaluation reveals that the uncommon data generated by our method is hard to be defended by the existing defense techniques with the average defense success rate reduced by 35\%.
Our results call for attention and necessity to generate more diverse data for evaluating quality assurance solutions of DL software.
\end{abstract}

\ccsdesc[500]{Software and its engineering~Software testing and debugging}
\ccsdesc[300]{Computing methodologies~Neural networks}

\keywords{Deep learning, uncertainty, adversarial attack, software testing}
	
\maketitle

\section{Introduction}

In company with the booming of available domain-specific big data and hardware acceleration, deep learning (DL) experienced big performance leap in the past few years, in achieving competitive performance in many cutting edge applications (\eg, image processing~\cite{ILSVRC15}, speech recognition~\cite{Hinton2012}, sentiment analysis~\cite{DO2019272},
e-commerce recommendation~\cite{Zhang:2019:DLB:3309872.3285029}, video game control~\cite{mnih2015humanlevel}). However, the state-of-the-art DL software still suffers from quality issues. 
A deep neural network (DNN) that achieves high prediction accuracy can still be vulnerable to adversarial examples (AEs)~\cite{biggio2013evasion}.
For example, an image recognition DL software can be easily fooled by pixel-level noises~\cite{one_pixel} or noises perceived in a physical-world situation~\cite{tian2018deeptest, DBLP:conf/cvpr/2018}.
The quality and reliability issues, without properly addressing or confining, could potentially hinder more widespread adoption of DL software especially in the applications with higher requirements of safety and security ~(\eg, autonomous driving, medical diagnosis).

The incorrect decision of a DL software can trace back to several typical sources and patterns (\eg, generalization capability issue, robustness issue~\cite{DBLP:conf/cav/KatzBDJK17}). Up to present, AEs remain to be one of the most notable types of DL defects, which reveals the quality and robustness issues of the DL software. Although the arms races of many recently proposed adversarial attack~\cite{Goodfellow2015,Seyed2016DeepFool,Kurakin2017adver,cw2017}, defense~\cite{gong2017adversarial, wang2018, labelsmooth, Papernot2016, xu2017feature, inputtransform, pixeldeflection} and testing~\cite{pei2017deepxplore,tian2018deeptest,deephunter19,deepstellar}
continuously escalate, most of these techniques are rather ad-hoc. 
Thus far, research efforts on understanding and interpretation of AEs and benign examples (BEs) are still at an early stage~\cite{NeurIPS2018_7998}.
Uncertainty provides a new perspective to characterize AEs and BEs, towards understanding and designing better quality assurance techniques for DL software, \eg, testing, adversarial attack/defense.

\begin{figure}[t]
    \centering
    \begin{subfigure}[b]{0.48\columnwidth}
    \centering
    \begin{minipage}{.9\linewidth}
    \includegraphics[width=\textwidth]{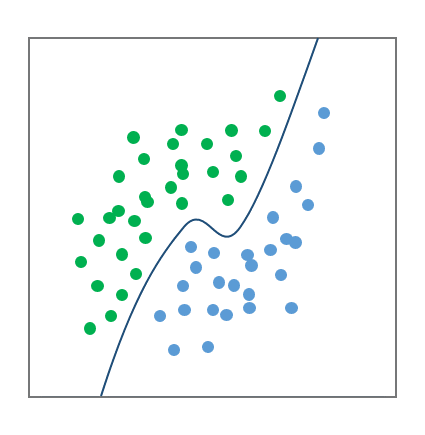}
    \end{minipage}
    \caption{Prediction Confidence}\label{fig:trace1}
    \end{subfigure}
    \begin{subfigure}[b]{0.48\columnwidth}
    \centering
    \begin{minipage}{.9\linewidth}
    \includegraphics[width=\textwidth]{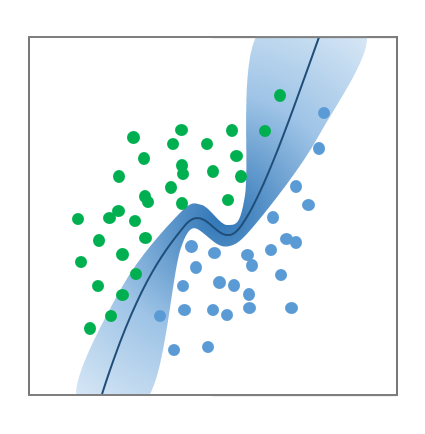}
    \end{minipage}
    \caption{Bayesian Uncertainty }\label{fig:trace2}
    \end{subfigure}
    \caption{Prediction Confidence and Bayesian Uncertainty.}
    \label{fig:uncertainty}
\end{figure}

To bridge this gap, in this paper, we study 4 state-of-the-art Bayesian uncertainty metrics (based on the statistical analysis of multi-shot executions) along with the one-shot execution metrics, \ie, prediction confidence score (PCS) of the DL software under analysis (see Fig.~\ref{fig:uncertainty} and definitions in \S~\ref{related}). 
We perform a large-scale comparative study to investigate the sensitivity/capability of these metrics in differentiating AEs and BEs, which are important indicators to characterize DL runtime behaviors.
We find that PCS and Variation Ratio in terms of original prediction (VRO) (\S~\ref{related}) are among the best candidates with such capability, and they are selected for further input data characterization.
Then, an obvious question arises: \emph{what is the relation between the AEs/BEs generated by the state-of-the-art adversarial attack/testing techniques and these uncertainty metrics}? In particular, \emph{do these AEs/BEs follow some patterns in terms of uncertainty}? 
Our in-depth analysis reveals that AEs/BEs generated by existing techniques~\cite{Goodfellow2015, Seyed2016DeepFool, bim, cw2017, tensorfuzz19, deephunter19} largely fall into two common patterns (that we refer as common input samples): (1) AEs tend to have low PCS and high VRO uncertainty, and (2) BEs often come with high PCS and low VRO. 

Based on the above observation, further questions naturally arise:
\emph{What do those uncommon inputs look like and can we generate them possibly and automatically?} and \emph{Would these uncommon data inputs differ from common inputs, \eg, are they even more challenging to be correctly handled by a DL software?}
To answer these questions, we propose a genetic algorithm (GA) based automated test generation technique that iteratively generates uncommon input samples guided by uncertainty metrics.
We implement the proposed technique as a tool named \tool(to \textbf{K}now the \textbf{U}n\textbf{K}nown), and demonstrate its effectiveness in generating uncommon inputs on a large benchmark, including three datasets MNIST, CIFAR10, ImageNet across four different DL model architectures \lenet, \nin, \res, \mobile.
In line with existing adversarial defense techniques, our comparative experiments against the state-of-the-art adversarial attack/testing techniques also reveal that the uncommon samples could be more challenging to be correctly handled by DL software in many cases.
Such uncommon samples represent a new type of hazard and potential defects to DL software, which so far lacks investigation and should draw further attention during future quality assurance solution design.

In summary, this paper investigates the following research questions with the support of a large-scale study:

\begin{itemize}[topsep=3pt,leftmargin=*]
    
    \item \textbf{RQ1:} What is the capability of the state-of-the-art uncertainty metrics in differentiating AEs and BEs?
    
    \item \textbf{RQ2:} Do the AEs and BEs generated by the state-of-the-art adversarial attack/testing techniques follow common uncertainty patterns? If so, what are such common patterns?
    
    \item \textbf{RQ3:} Is it possible to generate those uncommon data that is missed by the state-of-the-art attack/testing techniques? Is \tool useful in generating uncommon data?
    
    \item \textbf{RQ4:} To what extent are the uncommon samples defended by existing adversarial defense techniques compared with the common ones?
    
\end{itemize}

Through answering RQ1 and RQ2, we aim to characterize behaviors of DL software on AEs and BEs from the uncertainty perspective, and investigate whether some common patterns are followed by data inputs generated through current state-of-the-art adversarial attack/testing techniques. Our study confirms the existence of the common patterns for such generated data~(\ie, low PCS and high VRO for AEs, high PCS and low VRO for BEs). It also identifies new uncommon sample categories which are missed by existing techniques.

RQ3 and RQ4 focus on understanding the feasibility to obtain the uncommon samples (\ie, RQ3) and the impact of such samples on the quality and reliability of DL software (\ie, RQ4).
Our evaluation results confirm that the uncommon samples could be generated with proper testing guidance. Such uncommon samples, representing a new type of generated test data, could bypass a variety of adversarial defense techniques with higher success rates.
Thus, it is quite important to generate such uncommon inputs to reveal the hidden defects of DL software for vulnerability analysis and further enhancement, especially for safety- and security-critical scenarios.
We believe such uncommon data could be an important clue towards building trustworthy DL solutions, which should draw special attention for further quality assurance solution design.

The contributions of this paper are summarized as follows:

\begin{itemize}[leftmargin=*]
    \item We perform an empirical study on four state-of-the-art Bayesian uncertainty metrics and one prediction confidence metric to investigate their ability to differentiate BEs and AEs, \ie, data inputs that can/cannot be correctly handled by a DL software. Among these metrics, PCS and VRO outperform the others in achieving higher differentiating accuracy.
    
    \item We perform a systematic study on the BEs and AEs generated by six state-of-the-art adversarial attack/testing techniques (\ie,  FGSM~\cite{Goodfellow2015}, BIM~\cite{bim}, Deepfool~\cite{Seyed2016DeepFool}, C\&W~\cite{cw2017}, DeepHunter~\cite{deephunter19}, TensorFuzz~\cite{tensorfuzz19}) to identify potential uncertainty patterns in terms of PCS and VRO.
    Our results reveal that (1) AEs and BEs from existing techniques largely follow two common patterns while the other patterns (denoted as uncommon patterns) are largely missed by existing methods, and (2) the characteristics of AEs generated by the testing tools differ from those by the adversarial attack techniques.
    
    \item We propose a GA-based automated testing technique for DL software, implemented as a tool \tool, towards generating input samples with diverse uncertainty patterns. Especially, our evaluation on three datasets MNIST, CIFAR10, ImageNet across four different model architectures \lenet, \nin, \res, \mobile
    demonstrates its effectiveness in generating uncommon input samples that are largely missed by existing techniques.
    
    \item We further investigate how the adversarial defense techniques
    ~\cite{gong2017adversarial, wang2018, labelsmooth, Papernot2016, xu2017feature, inputtransform, pixeldeflection}
    react to the uncommon samples, in line with the samples generated by adversarial attack techniques.
    Our results indicate that the current defense techniques are often biased towards particular patterns of samples. 
    The generated uncommon samples can bypass such defense techniques with a high success rate, potentially causing severe threats to the quality and reliability of DL software. 
    For example, on the model \nin, the uncommon data achieve 97.5\% success rate on bypassing the mutation-based defense, while the common data only make it 5.5\%. 
    
\end{itemize}

\section{Preliminaries and overview}
\label{related}

\subsection{Deep Neural Networks}
\begin{Def}[Deep Neural Network]
A deep neural network (DNN) $M$ is a function $P=M(x)$ that maps an input $x$ to a predictive probability vector $P$. The output label of $M$ on input data $x$ is $L_{M}(x)=\operatorname*{argmax}_{i\in C} P[i]$ 
where $C$ is the set of classes and $P=M(x)$.
\end{Def}
In general, a DNN learns to extract features from the distribution of training data layer by layer, and provides the decision on each candidate class with some probabilistic confidence. A higher predictive probability value of a class often indicates higher prediction confidence on that decision label.

\begin{Def}[Benign Example]
A benign example (BE) $x$ with ground truth label $y$ is an input sample, such that the prediction decision of a DNN $M$ is consistent with the ground truth label:
$L_M(x) = y$.
\end{Def}

Benign examples refer to those inputs that could be correctly handled by a given DNN model $M$.

\begin{Def}[Adversarial Example]
An adversarial example (AE) is an input  $x^{\prime}$ similar to a benign example $x$ by adding some minor perturbation $\delta$, (i.e., $x^{\prime} = x + \delta$), but resulting in a different prediction decision of a DNN $M$ (i.e., $L_M(x^{\prime}) \neq L_M(x)$).
\end{Def}

Existing attack methods usually generate the AEs by manipulating the output of the \textit{logit} layer or \textit{softmax} layer of a DNN $M$, which gradually decreases in the output probability of ground truth label and increases in the probability of other labels. In this way, the prediction decision is shifted to other labels.
Based on direct observations on the prediction results of such AEs, we identify that a typical type of unreliable prediction is usually accompanied by two classes that have close probability confidence values.

\begin{Def}[Prediction Confidence Score]
Given a DNN $M$ and an input $x$, the prediction confidence score (PCS) of the input $x$ on $M$ is defined as 
\begin{equation*}
    PCS(x,M) = \max \limits_{i \in C} P[i] - \max \limits_{i \in C\setminus c^\ast} P[i]
\end{equation*}
where C is the set of classes, $c^\ast = L_{M}(x)$ and $P=M(x)$.
\end{Def}

Intuitively, \textit{PCS} depicts the probability difference between the two classes with the highest probabilities, which provides an uncertainty proxy from the aspect of distance to the geometric boundary. For an input $x$, the smaller the $PCS(x,M)$, the closer $x$ is to the decision boundary between the top-two classes.
As a result, it is more likely to cross the boundary with noise perturbations.
In the following sections, we use $PCS(X, M)$ to denote $\{PCS(x,M)|x\in X\}$, where $X$ is a set of inputs.

\subsection{Bayesian Uncertainty Measures}

Besides prediction confidence based metrics, Bayesian-based methods are recently proposed to estimate DNN's uncertainty through multi-shot execution analysis~\cite{Gal2016Uncertainty} (see Fig.~\ref{fig:uncertainty}).
From the principled Bayesian perspective, DNNs are not regarded as deterministic functions with fixed parameters. Instead, the parameters of DNNs are treated as random variables, which obey a \textit{prior} distribution $p(\omega)$. The posterior distribution $q(\omega | \mathcal{D})$ is then approximated given a training data set $\mathcal{D}$, based on which uncertainty estimates can be obtained. 
The relationship between the uncertainty representative of \textit{prediction confidence score} and \textit{Bayesian uncertainty estimates} is shown in Fig.~\ref{fig:uncertainty}. Intuitively, prediction confidence score reflects a distance abstraction to the fixed geometric boundary \wrt a point-estimate neural network whose parameters are fixed; while Bayesian method ensembles a set of networks whose weights follow some probability distribution.

However, due to the complexity of DNNs, it is often impractical to sample infinitely many weights from the distribution to perform the runtime execution. To this end, the state-of-the-art approach to obtain uncertainty estimates makes use of the dropout technique from multiple runs (\textit{Monte Carlo dropout}~\cite{gal2016dropout}).
Although dropout is originally proposed as a regularization method in the training process to avoid over-fitting,
in the context of uncertainty estimation, dropout is leveraged in the testing process, which samples weights from the distribution to obtain DNN instances.
As a result, it ensembles a number of neural networks with different weights. In each prediction execution, it randomly drops out some units in the DNN, which may cause different prediction results.
As a result, it allows to obtain uncertainty estimates efficiently and scales to real-world neural networks. Specifically, there are three commonly-used metrics to estimate uncertainty~\cite{Gal2016Uncertainty}, \ie, variation ratio, predictive entropy and mutual information.

\subsubsection{Variation Ratio}
Variation ratio measures the dispersion from the dominant class of the prediction (\ie, the predicted class with the highest frequency in multiple predictions). 
\begin{Def}[Variation Ratio]
Given a model $M$ and an input $x$, the \textit{variation ratio} (VR) of the input $x$ is defined as
\begin{equation*}
    VR(x, M) = 1 - \frac{\sum_{k\in\{1,\ldots,T\}}  L_{M_{d}^{k}}(x)= l_{max}}{T}
\end{equation*}
where $M_d$ is the model with dropout-enable and $L_{M_{d}^{k}}$ denotes the $k$-$th$ prediction result by $M_d$. $T$ is the total number of prediction execution by $M_d$. $l_{max}$ represents the dominant label from most of the $T$ predictions.
\end{Def}

Another variant of the \textit{variation ratio} in terms of the original prediction (the prediction of the model under analysis) is defined as follows:
\begin{Def}\label{def:vro}
Given a model $M$ and an input $x$, the \textit{variation ratio}  for original prediction (denoted as VRO) of the input $x$ is defined as
\begin{equation*}
    VRO(x, M) = 1 - \frac{\sum_{k\in\{1,\ldots,T\}}  L_{M_{d}^{k}}(x)= L_{M}(x)}{T}
\end{equation*}
where $L_{M}(x)$ is the prediction result from the original model $M$.
\end{Def}

Intuitively, \emph{VR} measures the general uncertainty of the decision with the highest frequency (\ie, whether most predictions agree with the same result) while \emph{VRO} represents the stability around the prediction mode of model $M$ (\ie, whether the majority predictions agree with the original result).
The higher the \emph{VR} or \emph{VRO} is, the more uncertain the prediction is.

\subsubsection{Predictive Entropy}
Predictive entropy originates from information theory and measures the average amount of information contained 
in a stochastic source of predictive output.
When all classes are predicted with equal probability in the form of a uniform distribution, 
the decision carries the most information, indicating high uncertainty. 
In contrast, when one of the classes is predicted with high probability value (\eg, 0.9), 
then the prediction is relatively certain. 

\begin{Def}
    Given all the predictive probability distributions $P_{d}^{k} = M_{d}^{k}(x)$, $k \in \{1, ..., T\}$ of an input $x$ across $T$ predictions
    on dropout enabling model $M_d$, the predictive entropy 
    is defined as
    \begin{equation*}
        PE(x, M) = - \sum_{i \in C}(\frac{1}{T}\sum_{k}P_{d}^{k}(i))\log (\frac{1}{T}\sum_{k}P_{d}^{k}(i))
    \end{equation*}
    where $P_{d}^{k}(i)$ denotes the probability value of a particular class $i$ on the $k$-$th$ prediction of model $M_d$.
\end{Def}

\subsubsection{Mutual Information}
Mutual information quantifies the amount of information 
obtained about one random variable 
through the observations of the other random variable. 
In the case of DNNs, the two random variables are prediction $L_M(x)$ 
and the posterior of the model parameters $\omega$, 
whose distribution is approximated through $T$ stochastic
forward passes of dropout enabled model $M_d$. We use $\omega_{k}$ to denote an instance of the model parameters sampled from the posterior distribution.
\begin{Def}
    Given the predictive probability distributions $P_{d}^{k} = M_{d}^{k}(x)$, $k \in \{1, ..., T\}$ of an input $x$ 
    in $T$ predictions, the mutual information of input $x$ is defined as 
    \begin{equation*}
        MI(x, M) = PE(x, M) + \frac{1}{T}\sum_{i, k} (P_{d}^{k}(i) \log P_{d}^{k}(i))
    \end{equation*}
    where $P_{d}^{k}(i)$ is used to approximate $P_{d}(i | \omega_{k})$  in a similar way to predictive entropy, based on the probability vectors obtained through different stochastic forward passes.
\end{Def}

\subsection{Overview}
\begin{figure*}
    \centering
    \includegraphics[width=1\linewidth]{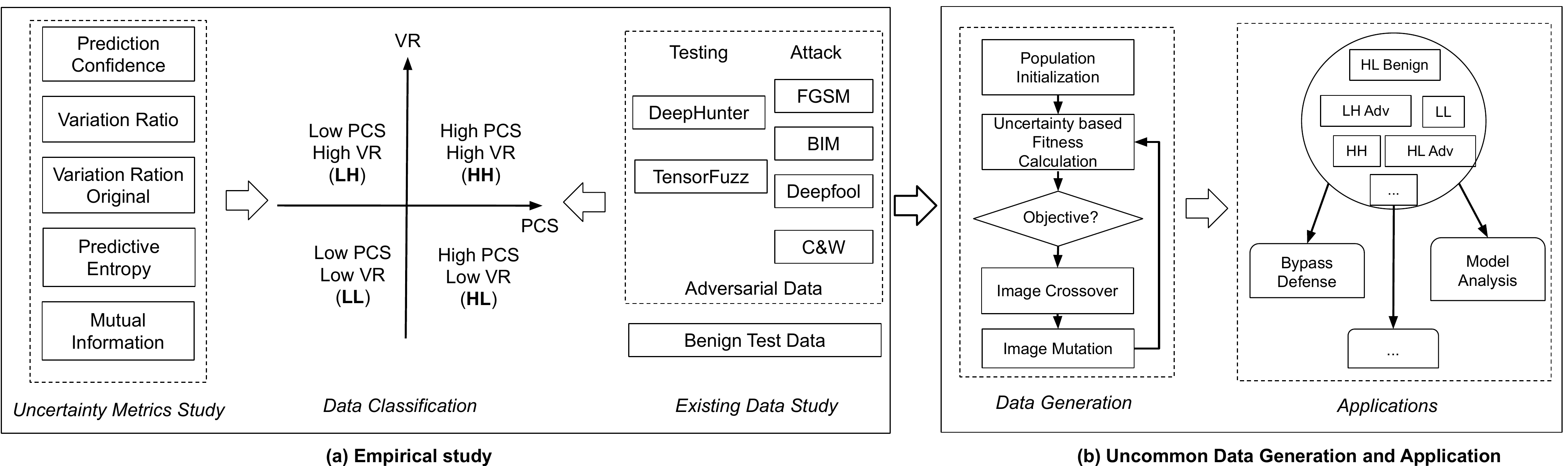}
    \caption{The overview of our study and its application.}
    \label{fig:overview}
\vspace{-2mm}
\end{figure*}

In this paper, we aim to understand the capability of different uncertainty portrayals, on which behaviors of AEs and BEs are further characterized.
Fig.~\ref{fig:overview} shows an overview of our work, summarized as two major components: (1) an empirical study about the uncertainty metrics and characteristics of the existing data inputs, and (2) the data generation algorithm that generates input samples with uncommon patterns and potential applications. 
Specifically, we first perform an empirical study towards understanding the capability of different uncertainty metrics on distinguishing AEs and BEs (\S~\ref{sec:metricstudy}). Then, we propose a way of categorizing AEs and BEs based on two perspectives: the \textit{prediction confidence score} from the perspective of the single-shot model execution and the \textit{uncertainty estimates} from the statistical perspective of multi-shot execution. Next, we study the uncertainty patterns of the BEs and AEs generated by the existing adversarial attack/testing tools (\S~\ref{sec:categori}).

From the empirical study, we find that the existing AEs and BEs follow specific uncertainty patterns. Then, We propose a genetic-algorithm-based approach to generate the uncommon data inputs (\S~\ref{sec:datagen}), whose uncertainty patterns are different from the patterns that existing data fall into. We identify and analyze the importance of data with diverse uncertainty patterns from a testing perspective. Finally, we evaluate the capability of such uncommon data in bypassing a variety of defense techniques (\S~\ref{sec:defenseevaluate}).

\section{Empirical Study}
\label{study}

In this section, we first perform a comparative study about the capability of different uncertainty metrics in distinguishing AEs and BEs. Then, we conducted a follow-up investigation of the characteristics of existing AEs/BEs from the uncertainty perspective.

\subsection{Subject Dataset and Data Preparation}

\subsubsection{Datasets}

\begin{table}[!t]
\caption{Subject datasets and DNN models.}
\small
\vspace{-2mm}
\begin{center}
\begin{tabular}{lcccl}
\toprule
 \textbf{Dataset} & \textbf{DNN Model} & \textbf{\#Neuron} & \textbf{\#Layer} & \textbf{Acc.(\%)}\tabularnewline

\toprule

 \textbf{MNIST} & LeNet-5 & 268 & 9 & 99.0\%\tabularnewline
\midrule 
\multirow{2}{*}{\textbf{CIFAR-10}} 
 & NIN & 1418 & 12 & 88.2\%\tabularnewline
  & ResNet-20 & 2,570 & 70 & 86.9\% \tabularnewline
\midrule 
\textbf{ImageNet}& MobileNet & 38,904 & 87 & 87.1\%*\tabularnewline

\bottomrule
\end{tabular}
\end{center}
\label{tab:model_summary_1}
\begin{center}
\scriptsize{* The reported top-5 test accuracy of pretrained DNN model in~\cite{mobilenets}}.
\end{center}
\vspace{-2mm}
\end{table}

We selected three popular publicly available datasets (\ie, MNIST~\cite{lecun1998}, CIFAR-10~\cite{cifar10}, and ImageNet~\cite{ILSVRC15}) as the evaluation subject datasets (see Table~\ref{tab:model_summary_1}). 

\textit{MNIST} is for handwritten digit image recognition, containing $60,000$ training data and $10,000$ test data, with a total number of $70,000$ data categorized in 10 classes~(\ie, handwritten digits from $0$ to $9$). Each MNIST image is of size $28 \times 28 \times 1$ with a single channel.

\textit{CIFAR-10} is a collection of images for general-purpose image classification, including $50,000$ training data and $10,000$ test data in $10$ different classes~(\eg, airplane, bird, cat and ship), with $6,000$ images per class. Each CIFAR-10 image is of size $32 \times 32 \times 3$ with three channels. 
The classification task of CIFAR-10 is generally difficult than that of MNIST due to the data size and complexity.

\textit{ImageNet} is a large-scale practice-sized dataset, which is used as the database for large-scale visual recognition challenge~(ILSVRC) towards general-purpose image classification. The complexity of ImageNet is characterized by a training set comprised of over 1 million images, together with a validation set comprised of $50,000$ images and a test set with $150,000$ images. Each image is of size $224 \times 224 \times 3$.

For each dataset, we study several popular DNN models used in previous work ~\cite{lecun1998, nin, deepres, mobilenets}, which achieve competitive test accuracy (Table~\ref{tab:model_summary_1}).

The DL models we study in this paper are all with convolutional architectures. However, our approach is generic and could be applied to other network architectures such as recurrent neural networks.
Our approach focuses on the uncertainty nature of DNN.
Whether the calculation methods of the uncertainty metrics are available determines the feasibility of applying our approach to DL models with other architectures.
Basically, the only requirement on the DL models is with a classification setting, thus the model output would be a probability distribution among a set of classes and the uncertainty metrics can be obtained.
In a word, our approach can be applied to DNN classification tasks, independent of the model architectures.

\subsubsection{Adversarial Example Generation Tools}
We chose four state-of-the-art adversarial attack tools, \ie, FGSM~\cite{Goodfellow2015} (Fast Gradient Sign Method), BIM~\cite{bim} (Basic Iterative Method), Deepfool~\cite{Seyed2016DeepFool} and C\&W~\cite{cw2017} attacks, to generate adversarial examples. 
Specifically, we used the existing Python tool-set \textsc{foolbox}~\cite{rauber2017foolbox} to perform these attacks and each attack is configured with the default setting.

We also selected two state-of-the-art automated testing tools for deep neuron networks, \ie, DeepHunter~\cite{deephunter19} and TensorFuzz~\cite{tensorfuzz19}, both of which adopt coverage-based fuzzing techniques. 
For DeepHunter, we generated AEs with the $k$-Multisection Neuron Coverage (KMNC) and Neuron Coverage (NC) as the guidance. Following the configuration in~\cite{deephunter19, pei2017deepxplore}, we set the parameter $k$ as 1,000 for KMNC and the threshold as 0.75 for NC. TensorFuzz is configured with the default setting used in~\cite{tensorfuzz19}.

\vspace{1mm}
\begin{table}[]
    \centering
    \small
    
    \caption{Number of adversarial examples generated by the testing tools and adversarial attacks.}
\begin{tabular}{c|ccc|c}
\toprule
Model & DH-KMNC & DH-NC & TensorFuzz & Adv\_attacks\tabularnewline
\toprule 
\lenet & 2,980 & 4,607 & 1,436 & 4$\times$9,000\tabularnewline
\nin & 5,715 & 9,110 & 3,848 & 4$\times$9,000\tabularnewline
\res & 6,960 & 9,249 & 2,465 & 4$\times$9,000\tabularnewline
\mobile & 3,314 & 20,541 & 13,999 & 4$\times$9,000\tabularnewline
\bottomrule 
\end{tabular}
    \label{tab:aegenerated}
\end{table}

\subsubsection{Data Preparation} \label{sec:dataprepare}
Overall, we prepared the following three sets of data: one set of benign examples,  one set of AEs generated by the attack methods, and one set of AEs generated by testing tools (see Table~\ref{tab:aegenerated}):
\begin{itemize}[leftmargin=*]

\item \textit{BenignData.} For each dataset, we randomly sampled $9,000$ test data, which can be correctly predicted by the models, as the benign dataset for each model.

\item \textit{AttackAdv.} For each input in \textit{BenignData}, we generated four types of AEs with the four attack methods, resulting in a total of $36,000$ AEs. (Column~\textit{Adv\_attacks} in Table~\ref{tab:aegenerated}). 

\item \textit{TestAdv.} For each dataset, we randomly sampled 500 benign examples as the initial seeds. Then, we ran DeepHunter and TensorFuzz for each model with 5,000 iterations. The columns \textit{DH-KMNC}, \textit{DH-NC}, and \textit{TensorFuzz} show the number of adversarial examples generated by DeepHunter with KMNC and NC guidance, and Tensorfuzz.
\end{itemize}

\subsection{RQ1: Empirical Study on Uncertainty Metrics}\label{sec:metricstudy}
The objective of RQ1 is to study the relationship between uncertainty metrics and adversarial examples. In particular, we analyze the effectiveness of uncertainty metrics in distinguishing AEs and BEs. 
We adopt AUC-ROC~\cite{roc} to evaluate the classification performance of each metric. We utilize the score as the evaluation criterion because it measures the performance without dependence on a pre-setting threshold.
Specifically, AUC-ROC is a performance measurement for classification problems at various threshold settings. ROC is short for receiver operating characteristic curve and AUC represents degree or measure of separability. It gives us detailed information on to what extent the evaluated model is capable of distinguishing between classes.
In our metric effectiveness evaluation, the higher the AUC-ROC score, the better a metric is at distinguishing AEs and BEs.

Table~\ref{tab:AUC-ROC} shows the AUC-ROC scores achieved on different metrics across different data. To be specific, we used $9,000$ BEs and $9,000$ AEs generated from each attack to calculate the AUC-ROC scores.
Overall, PCS achieves the best performance as it is a direct measure of the prediction confidence of the target model. 
Interestingly, we found that AEs often have low prediction confidence. 

From the Bayesian uncertainty perspective, we found that VRO achieves the best performance. 
On MNIST and CIFAR-10 datasets, all AUC-ROC scores are over 97\%. 
On ImageNet, the minimum AUC-ROC score is 77.47\% in differentiating BEs and AEs generated by Deepfool, while the scores on other attacks are all above 84\%. 
From the in-depth analysis, we found that VRO (refer to Definition~\ref{def:vro}) captures the difference between the prediction of the original model and multiple predictions of the randomized dropout-enabled model. 
On the other hand, other metrics represent the uncertainty based on the stability of the multi-shot predictions without considering the model under analysis. For example, given an input, suppose the prediction of the original model is incorrect and all predictions of the dropout-enabled model are correct, the values of VR and VRO would be 0 and 1, respectively. 
Intuitively, VRO shows that the model is quite uncertain but VR indicates that the model is very certain. In other words, VRO is more sensitive to capture the uncertain behavior of the target model compared with other uncertainty metrics.

\begin{tcolorbox}[size=title]
{\textbf{Answer to RQ1: PCS of the original model is often effective in distinguishing BEs and AEs generated by existing attacks. For the Bayesian uncertainty metrics, VRO is often more effective than others when comparing the prediction stability between the original model and multiple dropout-enabled predictions.
}} 
\end{tcolorbox}

\subsection{Characterizing Data Behavior}
PCS and Bayesian uncertainty depict the prediction results of a DNN from different angles.
In particular, existing studies demonstrate that high prediction confidence is not equal to low Bayesian uncertainty, and vice versa~\cite{Gal2016Uncertainty}. 

The prediction confidence (\ie, PCS) represents the confidence of a single-shot model execution while the Bayesian uncertainty is measured by the statistical results from multi-shot model executions. From RQ1, the results reveal that VRO stands out to capture the different behaviors of BEs and AEs compared with other Bayesian uncertainty metrics. 
Therefore, we adopt a two-dimensional metric, \ie, (PCS, VRO), to characterize the BEs and AEs on a specific DNN model.
Based on this, the data is classified into four patterns~(see Fig.~\ref{fig:overview} (a)): low \emph{PCS} and high \emph{VRO} (LH), high \emph{PCS} and high \emph{VRO} (HH), low \emph{PCS} and low \emph{VRO} (LL) and high \emph{PCS} and low \emph{VRO} (HL). The categorization provides a way to understand and analyze the behaviors of AEs and BEs.

\subsection{RQ2: Categorization of Existing Data} \label{sec:categori}

\begin{table}[]
    \centering
    \setlength\tabcolsep{4.0pt}
    \caption{The AUC-ROC scores of classification models with different metrics.}
\small
\vspace{-2mm}
\begin{tabular}{ccc|cccc}
\toprule 
Model & Attacks & PCS & VRO & VR & PE & MI\tabularnewline
\toprule
\multirow{4}{*}{LeNet-5} & BIM & \textbf{99.98\%} & \textbf{99.06\%} & 90.72\% & 83.56\% & 81.52\%\tabularnewline
 & C\&W & \textbf{100.00\%} & \textbf{99.08\%} & 90.23\% & 82.86\% & 81.61\%\tabularnewline
 & Deepfool & \textbf{99.44\%} & \textbf{98.31\%} & 93.47\% & 86.46\% & 84.78\%\tabularnewline
 & FGSM & \textbf{99.98\%} & \textbf{98.74\%} & 95.52\% & 90.09\% & 87.36\%\tabularnewline
\midrule
\multirow{4}{*}{NIN} & BIM & \textbf{99.95\%} & \textbf{99.46\%} & 88.57\% & 86.73\% & 86.95\%\tabularnewline
 & C\&W & \textbf{99.90\%} & \textbf{99.44\%} & 87.93\% & 85.99\% & 86.48\%\tabularnewline
 & Deepfool & \textbf{99.79\%} & \textbf{99.18\%} & 91.44\% & 88.64\% & 88.51\%\tabularnewline
 & FGSM & \textbf{99.43\%} & \textbf{98.80\%} & 93.97\% & 91.69\% & 91.54\%\tabularnewline
\midrule 
\multirow{4}{*}{ResNet-20} & BIM & \textbf{99.97\%} & \textbf{98.20\%} & 86.74\% & 84.62\% & 85.10\%\tabularnewline
 & C\&W & \textbf{99.88\%} & \textbf{98.28\%} & 85.87\% & 83.93\% & 84.59\%\tabularnewline
 & Deepfool & \textbf{99.80\%} & \textbf{97.85\%} & 88.02\% & 85.70\% & 86.23\%\tabularnewline
 & FGSM & \textbf{99.23\%} & \textbf{97.28\%} & 90.74\% & 88.25\% & 88.24\%\tabularnewline
\midrule 
\multirow{4}{*}{MobileNet} & BIM & \textbf{99.95\%} & \textbf{86.67\%} & 84.25\% & 68.77\% & 68.37\%\tabularnewline
 & C\&W & \textbf{96.80\%} & \textbf{84.49\%} & 82.73\% & 67.80\% & 67.35\%\tabularnewline
 & Deepfool & \textbf{79.36\%} & \textbf{77.47\%} & 77.13\% & 74.03\% & 72.32\%\tabularnewline
 & FGSM & \textbf{97.79\%} & \textbf{87.93\%} & 86.49\% & 72.75\% & 71.84\%\tabularnewline
\bottomrule
\end{tabular}

    \label{tab:AUC-ROC}
\end{table}

\begin{table*}[!t]
  \caption{Results (mean / variance) of benign \& adversarial data.}
  \small
  \label{map_result}
  \centering
  \begin{tabular}{ccc|cccc|ccc}
    \toprule
     \multirow{2}{*}{Model}  &\multirow{2}{*}{Metric} &\multirow{2}{*}{Benign} &\multicolumn{4}{c|}{AEs from Attacks}  &  \multicolumn{3}{c}{AEs from Testing tools} \tabularnewline
    \cmidrule(r){4-10}
    &  &   & BIM     & C\&W & Deepfool & FGSM&DH-KMNC&DH-NC&TensorFuzz \\
    \midrule
   \multirow{2}{*}{\lenet}  & PCS  & 0.990 / 0.004 & 0.018 / 0.001 & 0.002 / 8.865e-06 & {0.186 / 0.114} & 0.012 / 1e-4  &0.561/0.119& 0.592/0.112& 0.579/0.111  \\
   &VRO & 0.312 / 0.017 & 0.733 / 0.008 & 0.735 / 0.008 & 0.697 / 0.009 & 0.716 / 0.009&0.631/0.019&0.625/0.020&0.630/0.020\\
    \midrule
    \multirow{2}{*}{\nin}     & PCS       & 0.953 / 0.023 & 0.007 / 0.0004 & 0.013 / 0.0003 & 0.027 / 0.007 & 0.083 / 0.005 &0.571/0.113&0.608/0.110&0.384/ 0.093 \\
    &VRO &0.054 / 0.016 & 0.777 / 0.025 & 0.777 / 0.027 & 0.720 / 0.026 & 0.679 / 0.029&0.380/0.065&0.352/0.063&0.615/0.039\\
    \midrule
    \multirow{2}{*}{\res} & PCS & 0.945 / 0.026 & 0.005 / 0.0002 & 0.016 / 0.0005 & 0.023 / 0.006 & 0.096 / 0.006&0.548/0.108&0.574/0.109&0.392/0.093\\
    &VRO &0.072 / 0.023 &0.865 / 0.010 & 0.869 / 0.010 & 0.850 / 0.010  & 0.828 / 0.012&0.398/0.080&0.388/0.080&0.457/0.062\\
    \midrule
    \multirow{2}{*}{\mobile} & PCS & {0.788 / 0.082} & 0.002 / 1.598e-05 & 0.085 / 0.020 & 0.415 / 0.135 & 0.059 / 0.002 &0.601/0.160&0.659/0.119&0.614/0.126\\
    &VRO & 0.337 / 0.062 & 0.679 / 0.015 & {0.652 / 0.017} & 0.596 / 0.054  & 0.696 / 0.016&0.454/0.080&0.451/0.077&0.551/0.083\\

    \bottomrule
  \end{tabular}
  \vspace{-2mm}
\end{table*}
Based on the categorization method, we perform a study towards understanding the characteristics of BEs and AEs generated by adversarial attacks and testing tools. To compute the VRO, 
we follow the parameter configuration suggested in~\cite{gal2016dropout} and set $T$ (see Definition~\ref{def:vro}) as 50 for MNIST, and 100 for CIFAR-10 and ImageNet, respectively.

Table~\ref{map_result} summarizes the quantitative results of BEs and AEs, w.r.t. the two-dimensional metrics in four models. Note that the data in the columns \textit{Benign}, \textit{AEs from Attacks} and \textit{AEs from Testing tools} are from the three sets of data presented in \S~\ref{sec:dataprepare}. In each cell, the two values represent the mean and variance of the corresponding PCS and VRO metric results, respectively.

Overall, benign data mostly have high PCS and low VRO. The mean values of PCS for the four models are 0.99, 0.953, 0.945 and 0.788, respectively, while the mean values of VRO are 0.293, 0.054, 0.072 and 0.337. 
The results are mostly in line with our expectations, \ie, BEs are expected to be predicted by the model with high confidence and low uncertainty. 
In the case of MobileNet, the PCS is relatively smaller than others while the VRO is larger than that of other models. The reason is that MobileNet handles a more complex task, \ie, image classification for a large-scale dataset (ImageNet). It is more difficult to train a high-quality DNN model. MobileNet we used here is among the state-of-the-art model for image classification in ImageNet, whose top-5 accuracy is 87.1\%. 
This result indicates that BEs usually belong to the HL type.

For AEs generated by different attacks, the metric performance is mostly contrary to BEs, \ie, AEs generated by attacks usually come with low PCS and high VRO. 
Except for the AEs generated by Deepfool on ImageNet, all the mean values of PCS are rather low (\eg, almost all PCS values are below 0.1) and the mean values of VRO are relatively high (\eg, all VRO values are larger than 0.652). It indicates that AEs generated from state-of-the-art adversarial attacks usually fall into the category with low confidence and high uncertainty. From the results, we find that AEs usually belong to the LH type.

For the adversarial data generated from testing tools, we found that the obtained metrics are between BEs and AEs generated from the adversarial attacks. The mean values of PCS are smaller than the PCS results of BEs but often larger than the results of AEs from attacks. 
Conversely, the mean values of VRO are larger than the VRO results of BEs but smaller than the results of AEs from adversarial attacks. For example, consider the metric results of BEs, TensorFuzz AEs and C\&W AEs on ResNet-20, the PCS values are 0.945, 0.392 and 0.016, respectively, while the VRO values are 0.072, 0.457 and 0.869. 
Compared with the results of BEs, we can still reach a similar conclusion that AEs from testing tools also belong to the LH type.

Even so, the results of AEs from adversarial attacks and testing tools have some differences. This might be caused by their differences in the test generation methods. 
Current adversarial attacks usually adopt the gradient-based or optimization-based technique to gradually decrease the predictive probability or the logit value of the truth label until the decision result changes.
For example, when the probability of the truth label is reduced to 0.49 and the probability of another label becomes 0.51, an adversarial example is found and the attack stops. 
DeepHunter and TensorFuzz adopt the mutation-based technique to generate new tests. The random mutation can not guarantee that the predictive probability of the truth label is decreased gradually. For example, the probability may change from 0.99 to 0.10 by only one mutation, resulting in a higher PCS. 
In summary, although both adversarial attacks and testing tools can generate AEs, the behaviors of such generated AEs are different in terms of PCS and VRO. 
Therefore, this further confirms the difference between testing and adversarial attack, from the perspective of software engineering. Testing can not only simulate real-world scenarios to uncover potential issues of a DL software for deployment, but also generate more diverse data to capture the behaviors of the DNN systematically in the applied context.

Comparing the results of BEs and AEs, we also found that there also exists a tentative \textit{inverse correlation} between PCS and VRO. For example, the PCS of BEs is often large and the VRO is small. The PCS of AEs from adversarial attacks is often small while the VRO is large. For AEs generated from testing tools, the PCS tends to be larger and the VRO is smaller. 
It is reasonable since high prediction confidence usually reflects that the prediction is relatively certain.

\begin{tcolorbox}[size=title]
{\textbf{Answer to RQ2: BEs and AEs usually belong to the HL type and LH type, respectively. Compared with state-of-the-art adversarial attacks, testing tools to some extent generate different AEs.}}
\end{tcolorbox}

\section{Uncommon data generation}\label{sec:datagen}
The results of RQ2 reveal that BEs and AEs usually belong to the HL type and LH type, respectively. However, several questions still remain, \ie whether there exist: 1) data samples with high PCS and high VRO (\ie, the HH type), 2) data with low PCS and low VRO (\ie, the LL type), 3) BEs with low PCS and high VRO (\ie, the LH BEs) and 4) AEs with high PCS and low VRO (\ie, the HL AEs). 
These samples have the potential to uncover the unknown behaviors of DNN, which are largely missed by existing methods. To answer these questions, we developed a tool, \tool, to generate such uncommon data.\footnote{\scriptsize{We refer to these data as "uncommon" in the sense that they are rarely uncovered by existing techniques rather than that they rarely exist. Such data could occur widely in the real world.}}

As PCS and VRO on existing data usually follow an inverse correlation, it is non-trivial to generate the uncommon data. In fact, the test data generation of specific types could be a complicated optimization problem. In this paper, we leverage the Genetic Algorithm (GA)~\cite{genetic1995} to provide a solution.

Fig.~\ref{fig:overview} b) shows the workflow of our algorithm. The inputs of \tool include a seed $x$ (\ie, an initial image~\footnote{\scriptsize{In this paper, although we mainly focus on the image domain, the approach can also generalize to other domains.}}), a model $M$, the \emph{dropout} enabled model $M_d$ and a target type $c$. The output is a set of data samples that satisfy the \textit{objective}. 
We elaborate on the details of each step as follows.

\textbf{Population Initialization.} Given an input image, we first generate a set of images by randomly adding noise to it. 
In order to generate high-quality images (\ie, recognizable by human), we abandon the affine transformation (\eg, translation and rotation~\cite{deephunter19}) as the crossover may generate invalid images.
We use $L_\infty$ norm to constrain the allowable changes between the original seed and its generated counterpart.

\textbf{Objective and Fitness Calculation.} In each iteration, we check whether some generated images (in the population) satisfy the objective, which is specifically designed for each type based on PCS and VRO.
The test generation continues until some desirable data inputs are obtained.
To satisfy the objective, we design a set of \textit{piecewise fitness functions} to generate different types of uncommon data such that the higher the corresponding fitness value, the better the input. 

We use $X$ to denote the population, \ie, a set of images, and use $PCS(X,M)$ to denote $\{PCS(x,M)|x\in X\}$. 
Given the input $x$ and model $M$, the objectives and the fitness functions are defined as follows:
\begin{itemize} [leftmargin=*]
    \item For the\textit{ LL type}, the objective is $PCS(x,M)<p\wedge VRO(x,M)< v$, where $p$ and $v$ are configurable parameters, and the fitness function is:
\small
\begin{equation}
\label{equa:probmutationll}
Fit_{LL}(x) =
\left\{
\begin{array}{ll}
-PCS(x,M),  & min(PCS(X,M)) > p\ \\
(PCS(x,M) < p) - VRO(x,M) , &  \mathrm{otherwise} 
\end{array}
\right.
\end{equation}
\normalsize
If the minimum PCS of the population is larger than $p$, we use $-PCS(x,M)$ to decrease the PCS until there are some inputs whose PCSs are below $p$.
Due to the custom inverse correlation between PCS and VRO, VRO tends to increase when PCS decreases.
As for the fitness function in the other situation, ($PCS(x,M) < p$) aims to ensure that the PCS is still below $p$ while $-VRO(x,M)$ stands out for a smaller VRO.
\item  For the \textit{HH type}, the objective is $PCS(x,M)>p\wedge VRO(x,M)> v$, where $p$ and $v$ are configurable parameters, and the fitness function is:
\small
\begin{equation}
\label{equa:probmutationhh}
Fit_{HH}(x) =
\left\{
\begin{array}{ll}
PCS(x,M),  & max(PCS(X,M)) < p\ \\
(PCS(x,M) > p) + VRO(x,M) , &  \mathrm{otherwise} 
\end{array}
\right.
\end{equation}
\normalsize
If the maximum PCS of $X$ is smaller than $p$, we increase the PCS until some PCSs are larger than $p$. In addition, we ensure the PCSs is larger than $p$ ($PCS(x,M) > p$), and in the meantime attempt to increase VRO ($+VRO(x,M)$).

\item  To generate AEs which belong to the \textit{HL type}, the objective is set as: $PCS(x,M)>p\wedge VRO(x,M)< v \wedge x$ is an AE, where $p$ and $v$ are the configurable parameters, and the fitness function is:
\small
\begin{equation}
\label{equa:probmutationhl}
Fit_{HL}^A(x) =
\left\{
\begin{array}{ll}
-PCS(x,M),  &  \forall x'\in X, x' is~BE  \\
AE(x) + PCS(x,M), & \exists x'\in X, x' is~AE\\
&\wedge max(PCS(X,M))<p\\
AE(x) +(PCS(x,M)>p) & \mathrm{otherwise} \\
-VRO(x,M), & 

\end{array}
\right.
\end{equation}
\normalsize
where $AE(x)$ is 1 if $x$ is an AE. Otherwise, it is 0.

The generation of HL AEs is extremely challenging since HL is the typical feature of BEs.
To address the problem, we design a three-step approach. 
If all images in the population are BEs (step 1), we aim to generate AEs by decreasing the PCS, which is commonly used by state-of-the-art attacks.
Whenever any inputs become AEs and all PCSs become smaller than $p$ (step 2), we increase PCS but still keep the high priority of AE (\ie, AE has a high fitness value with the support of $AE(x)$). For example, if an AE becomes BE but achieves high PCS, its fitness value will decrease. In the last step, we set AE and high PCS as the high priority, then decrease VRO.

\item  To generate BEs that belong to \textit{LH type}, the objective is set as: $PCS(x,M)<p\wedge VRO(x,M) > v$ and $x$ is a BE. The fitness function is designed as follows:
\small
\begin{equation}
\label{equa:probmutationlh}
Fit_{LH}^B(x) =
\left\{
\begin{array}{ll}
BE(x)-PCS(x,M),  &  min(PCS(X,M)) >p  \\
BE(x) +(PCS(x,M)<p) & \mathrm{otherwise}  \\
+VR(x,M) & 
\end{array}
\right.
\end{equation}
\normalsize
where $BE(x)$ is 1 if $x$ is a BE. Otherwise, it is 0.

Similarly, the generation of LH BEs is also challenging as LH is a typical feature of AEs. In the first step, if all PCSs are larger than $p$, we first decrease the PCS but keep the high priority of BE. In the second step, when there are some BEs with low PCSs, we increase the VRO but keep the high priority of BE and low PCS.

\end{itemize}
\textbf{Crossover and Mutation.} For the crossover, we adopt the tournament selection strategy to select two tournaments. From each tournament, we select one image, which has the largest fitness value. The two selected images are used to perform the crossover by randomly exchanging the corresponding pixels. After the crossover, each image is randomly mutated by adding white noise, to increase the diversity of the population. The test generation continues until the objective is satisfied or the given computation resource~(\eg, time limit) exhausts.

\section{Evaluation}
\label{eval}
We implemented the proposed test generation tool, \tool, in Python based on  Keras~\cite{keras}~(2.2.4) with TensorFlow~\cite{tensorflow}~(1.12.0) as backend. In this section, we aim to evaluate the usefulness of \tool in generating uncommon data (\textbf{RQ3}) and the effectiveness of these data in bypassing the defense techniques (\textbf{RQ4}).
All the experiments were run on a server with the Ubuntu 16.04 system with 28-core 2.0GHz Xeon CPU, 196 GB RAM and 4 NVIDIA Tesla V100 16G GPUs.

\subsection{RQ3: Usefulness of Test Data Generation}
\begin{figure}[!t]
     \includegraphics[width=0.8\linewidth]{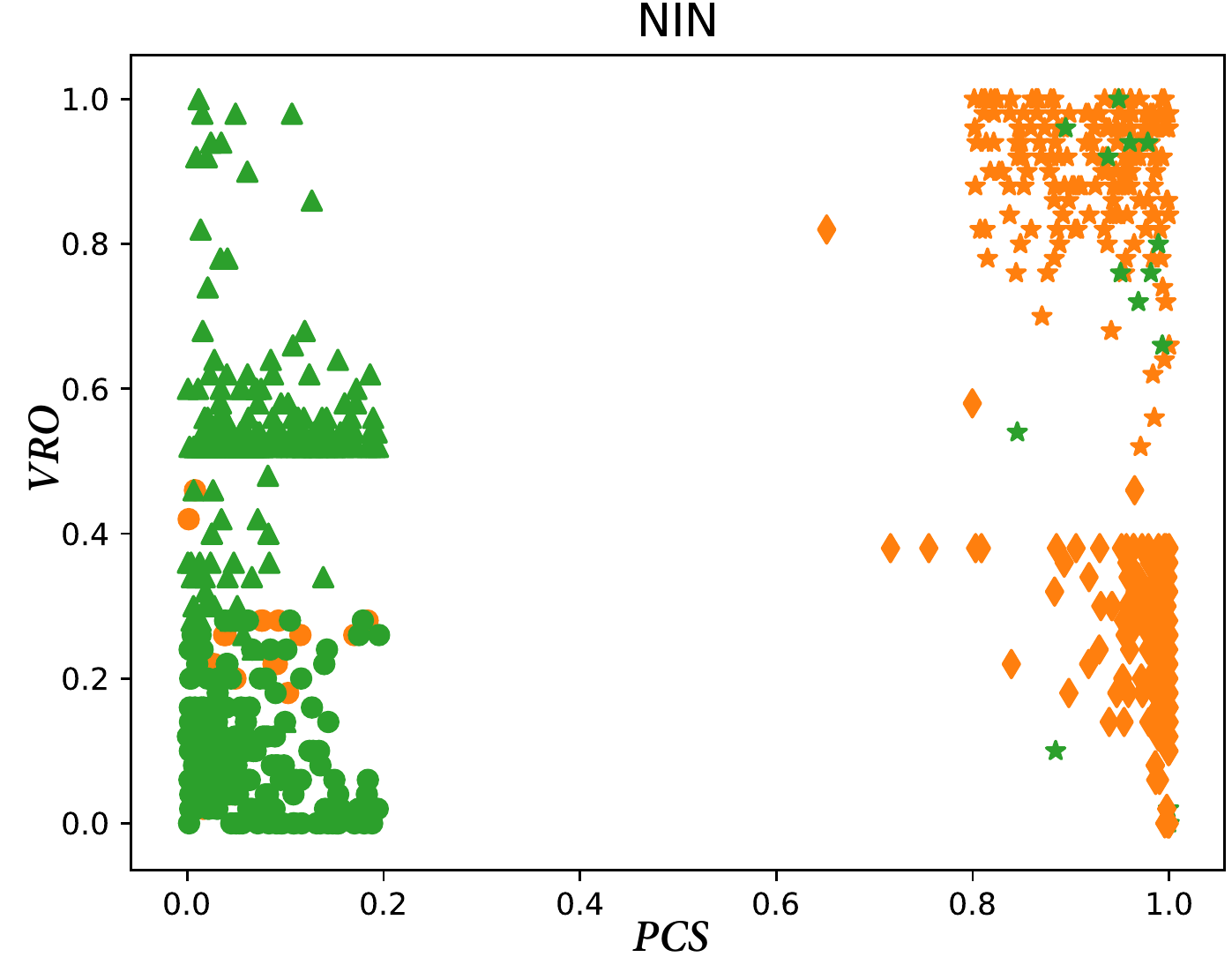}
   \captionof{figure}{
   The distribution of different types of data generated on \nin by \tool. The orange color represents AEs, while green color represents BEs. Star, circle, triangle and diamond represent the data of HH, LL, LH and HL, respectively. 
   }
   \label{fig:distribute}
\end{figure}

\textbf{Setting.} We adopt \tool to generate different types of uncommon data on four widely used DL models -- \lenet, \nin, \res and \mobile. In the genetic algorithm, the size of the population is set as 100, the crossover rate is set as 0.5 and the mutation rate is 0.005. For the mutation process, the radius of $L_\infty$ is set as $0.3$. For each dataset, we randomly select 200 BEs as the initial seeds. For each seed, we generate four types of uncommon data. The maximum number of the iterations in the genetic algorithm is set to 50.

\noindent\textbf{Threshold Selection.} To perform the categorization, we need to set the upper bound for the low PCS/VRO, the lower bound for the high PCS/VRO, \ie, the configuration values of the parameters $p$ and $v$ in the objective and fitness functions. Actually, in Table~\ref{map_result}, the results of BEs (HL) and AEs (LH) generated from adversarial attacks are the extreme cases, which can act as the guidance to select the thresholds. 
\begin{itemize}[leftmargin=*]

\item \textit{High PCS.} For the lower bound of high PCS, we set it as 0.7 as all high PCSs from BEs in the four models are above 0.7. Specifically, the minimum value of high PCS of BEs is 0.788 in \mobile. Intuitively, if the PCS of a data sample is above 0.7, we regard it as a data sample with \textit{high PCS}.

\item \textit{Low VRO.} For the upper bound of the low VRO, we set the value as 0.4 since the low VROs of BEs in the four models are below 0.4 (\eg, 0.312 and 0.337 for \lenet and \mobile, respectively). If the VRO of a data sample is below 0.4, we categorize it into \textit{low VRO} type.

\item \textit{Low PCS.} For the upper bound of the low PCS, most of the low PCSs of AEs in Table~\ref{map_result} are below 0.1 while the PCS of samples generated by Deepfool for \mobile has a larger value 0.415. We made a compromise and set the value as 0.3. If the PCS of a data sample is below 0.3, we regard that it falls into \textit{low PCS} category.

\item \textit{High VRO.} For the lower bound of high VRO, we set it as 0.6 because almost all of the high VROs of AEs in the four models are larger than 0.6 (one exception is 0.596 in \mobile). If the VRO is above 0.6, we regard it as \textit{high VRO}.
\end{itemize}

\begin{table*}[!t]

\caption{Results of different types of data generated by \tool.}
    \centering
\small
\vspace{-2mm}

\begin{tabular}{cc|cc|cc|c|c}
\toprule
\multirow{2}{*}{Type Objective} & & \multicolumn{2}{c|}{\textbf{LL} (<0.3/<0.4)} & \multicolumn{2}{c|}{\textbf{HH} (>0.7/>0.6)} & \textbf{LH} (<0.3/>0.6) & \textbf{HL} (>0.7/<0.4) \tabularnewline
\cline{3-8}
 &  & Ben & Adv & Ben & Adv & Ben & Adv\tabularnewline
\toprule
\multirow{2}{*}{LeNet-5} & Total & 124 & 18 & 130 & 46 & 172 & 1\tabularnewline
 & PCS/VRO & 0.067/0.324 & 0.030/0.343 & 0.947/0.720 & 0.903/0.803 & 0.082/0.609 & 0.99/0.339\tabularnewline
\hline 
\multirow{2}{*}{NIN} & Total & 176 & 22 & 10 & 184 & 58 & 165\tabularnewline
 & PCS/VRO & 0.065/0.095 & 0.065/0.204 & 0.960/0.846 & 0.922/0.910 & 0.073/0.666 & 0.981/0.245\tabularnewline
\hline 
\multirow{2}{*}{ResNet-20} & Total & 168 & 32 & 11 & 181 & 67 & 93\tabularnewline
 & PCS/VRO & 0.065/0.105 & 0.064/0.224 & 0.960/0.851 & 0.921/0.910 & 0.058/0.655 & 0.978/0.253\tabularnewline
\hline 
\multirow{2}{*}{MobileNet} & Total & 83 & 5 & 70 & 11 & 152 & 74\tabularnewline
 & PCS/VRO & 0.234/0.348 & 0.192/0.348 & 0.845/0.714 & 0.784/0.785 & 0.092/0.719 & 0.915/0.328\tabularnewline
\bottomrule 
\end{tabular}
\vspace{-3mm}
\label{tab:generation}
\end{table*}

\textbf{Results.} Fig.~\ref{fig:distribute} depicts the distribution of the 200 generated data on the two-dimension plane (due to page limit, the results of other models are put on our website~\cite{ourweb}). 
Note that there are some seeds from which we failed to generate the uncommon data satisfying the objective. For each of these seeds, we plot the best result from the population (\eg, for HH type, we select the data, which has the maximum of the sum of PCS and VRO).
The results show that \tool enables to generate inputs with diverse uncertainty patterns.

Table~\ref{tab:generation} shows the number of uncommon data inputs that satisfy the objective and the mean PCS and VRO value. Row \textit{Type Objective} shows the objective setting for each uncommon uncertainty pattern in terms of upper and lower bound of PCS and VRO. For each model, Row \textit{Total} and \textit{PCS/VRO} give the total number of uncommon data generated for each type and the mean value of PCS and VRO, respectively. For LL and HH types, the results of generated AEs and BEs are shown separately (\ie, Column \textit{Ben} and Column \textit{Adv}). 

The results demonstrate that \tool is effective in generating LL and HH data inputs that are rarely covered by existing methods.
For example, for \nin model, \tool generated 198 (99\%) LL data in total, of which 176 data are BEs and 22 data are AEs. For \res, \tool generated the LL data and HH data for all seeds (\ie, 200). From the quantitative result of LL data, we could find that LL data tend to be BEs (\eg, the number of LL BEs is much larger than the number of LL AEs). Considering the natural BEs usually belong to HL (refer to Table~\ref{map_result}), it indicates that low VRO is a better metric to represent the characteristics of BEs. For the HH data, there is no such obvious trend. In particular, for \lenet and \mobile, the number of HH BEs is larger than the number of HH AEs. However, the case in \nin and \res is on the contrary.

For LH BEs and HL AEs, it is more challenging to generate them since they are completely opposite to the characteristics of the common data. The results show that \tool is useful to generate such uncommon data. For example, we generated 172  (86\%), 58 (29\%), 67 (33.5\%) and 152 (76\%) LH BEs for \lenet, \nin, \res and \mobile, respectively. For HL AEs, we found that \tool only generated one HL AE for \lenet. For other models, \tool generated 165 (82.5\%), 93 (46.5\%) and 74 (37\%) HL AEs, respectively. The results indicate that generating LH BEs and HL AEs is more difficult and \tool can still generate them for a part of seeds.

We can see that the PCS and VRO in Table~\ref{tab:generation} are consistent with those in Table~\ref{map_result}. For example, although we set the lower bound of high PCS as 0.7 in the fitness functions and objectives, \tool still generated very high PCS for HH and HL data. The mean value of PCS is larger than 0.9 in \lenet, \nin, and \res, which is very consistent with the high PCS of BEs in Table~\ref{map_result}.
Even we set the upper bound of low VRO as 0.4, \tool still generated data with pretty low VRO, \eg, for \nin and \res, the VROs of LL data are 0.095 and 0.105, respectively.

Comparing the results among the four models, we could find that the difficulty in generating uncommon data varies for different models. For example, \tool is effective in generating LL and HH data for \lenet, \nin, and \res, but only generated 88 (44\%) and 81 (40.5\%) for \mobile. For \lenet and \mobile, it is easier to generate LH BEs than HL AEs but the case in \nin and \res is on the contrary. These differences show that the uncommon data generated through \tool can be used to characterize different behaviors of the models.

\begin{tcolorbox}[size=title]
{\textbf{Answer to RQ3: \tool is useful for generating different types of uncommon data. The HL BEs and HL AEs are often more difficult to be generated.}}
\end{tcolorbox}

\subsection{RQ4: Evaluation on Defense Techniques}\label{sec:defenseevaluate}
To demonstrate the usefulness of the generated uncommon data in Table~\ref{tab:generation}, this experiment intends to study whether the data can bypass the existing defense techniques.

\textbf{Setting.} 
Since different defense techniques are proposed on different subject datasets, we selected popular techniques based on the datasets. 
For MNIST and CIFAR10 dataset, we selected the following defense techniques:  binary activation classifier~\cite{gong2017adversarial}, mutation-based adversarial attack detection~\cite{wang2018}, defensive distillation~\cite{Papernot2016}, label smoothing~\cite{labelsmooth}, 
and feature squeezing~\cite{xu2017feature}.
For ImageNet, we selected the mutation-based adversarial attack detection~\cite{wang2018}, input transformations~\cite{inputtransform} and pixel deflection~\cite{pixeldeflection}. Due to the space limit, we put the details about the configuration and the introduction of each defense technique on our website~\cite{ourweb}. 

To validate the performance of the defense techniques, we selected 1) the common data including 9,000 BEs and 9,000 AEs generated from the existing adversarial attacks (\S~\ref{sec:dataprepare}), and 2) the uncommon data from Table~\ref{tab:generation}. We use the success rate to evaluate the capability of the defense technique, \ie, divide the number of BEs/AEs, which can be correctly identified, by the total number.

\textbf{Results.} Table~\ref{tab:Detection} and~\ref{tab:mobiledec} show the results about the performance of the defense techniques on four models. Column/Row \textit{Comm} represents the success rate on existing common data. The results show that the defense techniques are very effective in identifying BEs and AEs of the common data, especially on the smaller models. For example, except for that the success rate of feature squeezing is 88.1\% on \res, the success rates of other defense techniques are above 90\% on \nin, \res, and \lenet. In particular, the success rate is above 97\% on \lenet. For the larger model \mobile, the success rate is relatively low as it is more difficult to perform the defense for the complex model.

Column/Row \textit{UnCo} represents the success rate on the uncommon data generated from \tool.
The overall result shows that the existing defense techniques perform poorly on the uncommon data we generated. We could find that the success rate of the \textit{binary classifier} and \textit{mutation-based} detection are reduced a lot. For example, on \nin, the success rate is reduced to 25.5\% and 5.5\%, respectively. The reason is that these two techniques mainly depend on the PCS and VRO characteristics for detection. Specifically, \textit{binary classifier} is trained with the value of \textit{logit} layer that is closely related to prediction confidence and the label change ratio in \textit{mutation-based} detection is similar to VRO. The uncommon data is very different from the common data \wrt these two metrics. As a result, if they perform well on common data, the success rate on the uncommon data could be low.

For other defense techniques, the reduction in success rate appears smaller than that of \textit{binary classifier} and \textit{mutation-based detection}. 
For example, the success rates drop to 78.3\% and 76.2\% for \textit{defensive distillation} and \textit{label smoothing} on \nin.
The reason is that a new model is retrained with these defense techniques, while the attacks are generated regarding to the original one, making it a more challenging transfer attack scenario.
For example, \textit{defensive distillation} retrains a more robust model by reducing the gradients.
In this case, some of the data, which are uncommon for original model, become common data \wrt the retrained model, because of some weight variation.
However, it still can be seen from the results that the uncommon data reveal stronger transferability.

We also found that the uncommon data on \lenet is not effective on \textit{defensive distillation} and \textit{label smoothing}. We performed an investigation and found that it may be caused by the following reasons. 1) Most of the data become common in the new model. It confirms the usefulness of uncommon data in characterizing the different behaviors of multiple models. 2) Most of the uncommon data generated on \lenet are BEs (see Table~\ref{tab:generation}). The success rate is reduced to 90.8\% and 79.9\% if we only use the uncommon AEs.

\begin{table}[t]
    \centering
    \small
    \setlength\tabcolsep{4.0pt}
    \caption{Success rate of the defense techniques on the generated data for \nin, \res and \lenet.}

\begin{tabular}{ccccccc}
\toprule 
 & \multicolumn{2}{c}{NIN} & \multicolumn{2}{c}{ResNet-20} & \multicolumn{2}{c}{LeNet5}\tabularnewline
\cline{2-7} 
 & Comm & UnCo & Comm & UnCo & Comm & UnCo\tabularnewline
\midrule
\#Data & 18,000 & 615 & 18,000 & 552 & 18,000 & 491\tabularnewline
\midrule
binary classifier & 0.944 & 0.255 & 0.958 & \textbf{0.159} & 0.986 & 0.303\tabularnewline
mutation-based & 0.975 & \textbf{0.055} & 0.985 & \textbf{0.101} & 0.97 & 0.390\tabularnewline

distillation & 0.93 & 0.786 & 0.913 & \textbf{0.773} & 0.985 & 0.963\tabularnewline
label smoothing & 0.936 & 0.762 & 0.921 & \textbf{0.748} & 0.981 & 0.934\tabularnewline
feature squezzing & 0.905 & 0.663 & 0.881 & 0.637 & 0.973 & \textbf{0.327}\tabularnewline
\bottomrule 
\end{tabular}
    \label{tab:Detection}
\end{table}

\begin{tcolorbox}[size=title]
{\textbf{Answer to RQ4: The uncommon data inputs are not well defended by existing defense techniques while the common data are relatively easier to be defended. In particular, the \textit{binary classifier} and \textit{mutation-based detection} approaches are less useful in defending the uncommon data inputs (\eg, with only 5\% and 10\% on \nin and \res, respectively).
}}
\end{tcolorbox}

\section{Threat to Validity}
The selection of the subject datasets and DNN models could be a threat to validity. We try to counter this by using three publically available datasets with diverse scales, and popular pre-trained DNN models that achieve competitive performance. 

The selection of the thresholds for the categorization may affect the results of Table~\ref{tab:generation}. 
We carefully select the thresholds based on the results of Table~\ref{map_result}. Furthermore, the results of Table~\ref{tab:generation} are relatively consistent with the results of Table~\ref{map_result}, indicating that the selection basically does not affect the results of Table~\ref{tab:generation}.

A further threat would be the randomness factors for computing the VRO (\ie, configuration parameter $T$ in \S~\ref{def:vro}). The previous work~\cite{Carlini2017bypass} found that the result is not sensitive to the choice of $T$ as long as $T$ is greater than 20 in CIFAR-10 and MNIST. We follow the configuration in the existing paper~\cite{gal2016dropout}, \ie, 50, 100, 100 for MNIST, CIFAR-10 and ImageNet. In addition, we tested that the values are sufficient for computing the stable VRO.

\section{Related work}

In this section, we summarize the most relevant work to ours.

\noindent \textbf{Attack and defense.} Ever since the demonstration of deep learning models being vulnerable to even a small perturbation of input data~\cite{szegedy2013intriguing}, a sequence of attack techniques have been developed on the strand. To date, multiple types of attacks including \emph{FGSM}~\cite{Goodfellow2015}, \emph{JSMA}~\cite{papernot2016SP}, \emph{BIM}~\cite{Kurakin2017adver}, \emph{DeepFool}~\cite{Seyed2016DeepFool}, \emph{C\&W}~\cite{cw2017} have been proposed; a parallel research focus on improving the robustness of deep learning models. \citet{Goodfellow2015} presented a method of introducing nonlinear model families into the training process. Defensive distillation was introduced to reduce the effectiveness of adversarial samples~\cite{Papernot2016}, which was then broken by \emph{C\&W} attack \cite{cw2017}. 
Meanwhile, a set of recent defense techniques was surveyed and shown that all could be defeated by constructing new loss functions \cite{Carlini2017bypass}.  
A more recent work~\cite{Madry18} exploited the framework of robust optimization for network adversarial training to resist a wide range of attacks.
Besides dealing with datasets like MNIST and CIFAR10, 
defense techniques~\cite{inputtransform, pixeldeflection} were also proposed to handle real-world large-scale datasets like ImageNet. 
However, it still lacks a study on the characteristics of BEs and AEs generated through these methods, which we attempt to attain from the uncertainty perspective.

\noindent \textbf{Uncertainty measures.} In general, a deep learning model is trained with a dataset and results in a set of fixed parameters, which further sets up a deterministic function mapping an input to a probability distribution. 
Bayesian approach, however, does not view deep learning models as deterministic functions; instead, they treat the parameters as random variables \cite{mackay1992}.
As a representative work to solve the scalability problem of obtaining uncertainty measures, \cite{gal2016dropout} proposed a dropout-based solution, which allows us to calculate uncertainty estimates of existing deep learning models with a good trade-off between uncertainty quality and computational complexity. 
Existing research on uncertainty measure applications mainly
focuses on the adversary detection problem. 
For example, \cite{feinman2017detecting} used both density estimates
and Bayesian uncertainty estimates to learn a regression model for adversarial example detection.
Further, an empirical study on two types of uncertainty measures, predictive entropy and mutual information, was proposed to understand the effectiveness of them for detection~\cite{smith2018understanding}. 
However, in our work, we perform a comparative study on both single-shot and multi-shot execution uncertainty estimates to dig out the uncertainty patterns that existing AEs/BEs followed.

\begin{table}[t]
    \centering
\small
\caption{Success rate of the defense techniques on the generated data for \mobile.}

\begin{tabular}{ccccc}
\toprule 
 & \#Data & mutation-based & input trans & deflection\tabularnewline
\midrule
Comm & 18,000 & 0.865 & 0.702 & 0.782\tabularnewline
UnCo & 395 & \textbf{0.549 }& 0.660 & 0.688\tabularnewline
\bottomrule
\end{tabular}
    \label{tab:mobiledec}
\end{table}

\noindent \textbf{Testing and debugging.}
Researchers have attempted to leverage decades of advances in the software engineering community to seek for solutions towards more secure and robust DL systems and have developed a set of fruitful results.
These approaches share a different spirit from those in the DL community, and, for the first time, have been evidenced to be unique and promising without studies.
Testing criteria come out as the first research focus.
A series of measurements have been adapted to evaluate the quality of DL test dataset, including neuron coverage~\cite{pei2017deepxplore}, multi-granularity coverage criteria~\cite{ma2018deepgauge}, MC/DC test criteria~\cite{sun2018testing},
combinatorial testing criteria~\cite{2018arXiv180607723M}
, surprise adequacy~\cite{DBLP:journals/corr/abs-1808-08444} and uncertainty-based metrics~\cite{ma2019test}. 
Classic testing methodologies have also been incorporated for DNN testing, including differential testing~\cite{pei2017deepxplore}, coverage-guided testing~\cite{deephunter19,tensorfuzz19,tian2018deeptest},
mutation testing~\cite{8539073} and concolic testing~\cite{sun2018}.
Some advanced test generation methods~\cite{deeproad_ASE18, deepstellar, diffchaser} have also been proposed to achieve better testing for different applications.
Similar to samples generated by adversarial attacks, it still lacks a study on the relationship of samples generated by testing techniques and uncertainty. In addition, the results of this paper confirm the difference between testing and adversarial attack, in obtaining samples with different uncertainty behaviors.

Some recent efforts have been made to debug DL models~\cite{ma2018mode}, and to study DL program bugs~\cite{ZhangISSTA}, library bugs~\cite{Pham:2019:CCO:3339505.3339633} and DL software bugs across different frameworks and platforms~\cite{guo2019empirical}.
The results of this paper provide a new angle to characterize DL model defects, which could be useful for other quality assurance activities besides testing.

\section{Conclusion and Future Work}
This paper performed an empirical study to characterize the data inputs from the perspective of uncertainty. We first presented an empirical study on the capability of uncertainty metrics in
differentiating AEs and BEs. Then, we performed a systematic study of the characteristics of BEs and AEs generated by existing attack/testing methods in terms of uncertainty metrics. The results reveal that existing BEs and AEs largely fall into two uncertainty patterns in terms of PCS and VRO. 
Based on the investigation results, we proposed a GA-based automated test generation technique to generate data with more diverse uncertainty patterns, especially those uncommon samples.
The results demonstrated the usefulness of the generated data 
in bypassing defense techniques. In future, we plan to perform more in-depth investigations on the application of the uncommon data towards robustness enhancement.
We believe further understanding of these uncommon data is crucial for building reliable and trustworthy DL solutions.

\begin{acks}
We thank the anonymous reviewers for their comprehensive feedback. This research was supported (in part) by the National Research Foundation, Prime Ministers Office, Singapore under its National Cybersecurity R\&D Program (Award No.NRF2018NCR-NCR005-0001), National Satellite of Excellence in Trustworthy Software System (Award No.NRF2018NCR-NSOE003-0001);
JSPS KAKENHI Grant No.19K24348, 19H04086, 18H04097, Qdai-jump Research Program No.01277; the National Natural Science Foundation of China under grant No.61772038, 61532019, and the Guangdong Science and Technology Department (Grant No.2018B010107004).
We also gratefully acknowledge the support of NVIDIA AI Tech Center (NVAITC) to our research.
\end{acks}

\clearpage
\bibliographystyle{ACM-Reference-Format}
\bibliography{reference}
\end{document}